\documentclass[a4paper,10pt]{article}

\usepackage{graphicx}
\usepackage{wallpaper}
\ThisULCornerWallPaper{1}{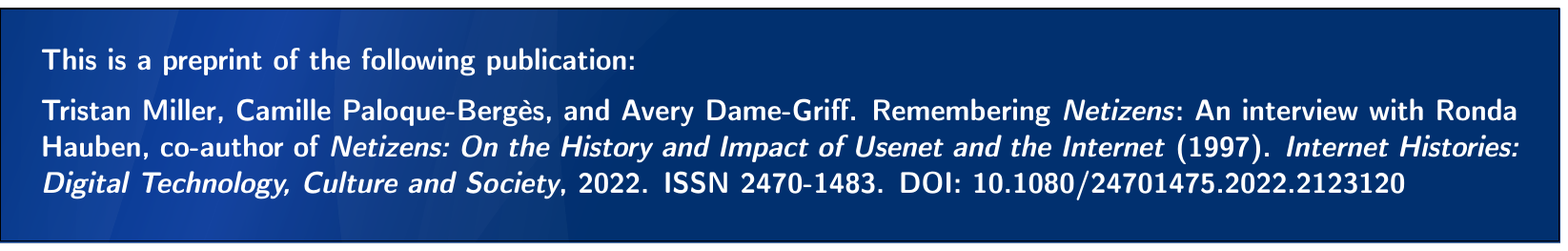}

\usepackage[T1]{fontenc}
\usepackage{CJKutf8}
\usepackage[american,british]{babel}
\usepackage{microtype}
\usepackage{parskip}
\usepackage[hyphens]{url}
\usepackage[breaklinks,colorlinks]{hyperref}
\usepackage{csquotes}
\usepackage{xpatch}
\usepackage[style=apa,
backend=biber,
uniquename=false,
sortcites=false]{biblatex}
\newcommand\iquestion[1]{\medskip\foreignlanguage{british}{\textbf{#1}}}
\hypersetup{
  linkcolor=blue,
  citecolor=blue,
  urlcolor=blue,
  pdftitle={\texorpdfstring{Remembering \emph{Netizens}:\\
An Interview with Ronda Hauben, co-author of \emph{Netizens: On the
History and Impact of Usenet and the Internet}
(1997)}{Remembering Netizens: An Interview with Ronda Hauben, co-author of Netizens: On the History and Impact of Usenet and the Internet (1997)}},
  pdfauthor={Tristan Miller, Camille Paloque-Bergès, and Avery Dame-Griff},
}

\begin{document}

  \title{\texorpdfstring{Remembering \emph{Netizens}:\\
An Interview with Ronda Hauben, co-author of \emph{Netizens: On the
History and Impact of Usenet and the Internet}
(1997)}{Remembering Netizens: An Interview with Ronda Hauben, co-author of Netizens: On the History and Impact of Usenet and the Internet (1997)}}
\author{Tristan Miller\\
Austrian Research Institute for Artificial Intelligence\\
Vienna, Austria
\and
Camille Paloque-Bergès\\
HT2S, Conservatoire national des arts et métiers\\
Paris, France
\and
Avery Dame-Griff\\
Women's and Gender Studies, Gonzaga University\\
Spokane, WA, USA}
\date{}

\maketitle
\thispagestyle{empty}

\begin{abstract}
  \emph{Netizens}, Michael and Ronda Hauben's foundational treatise on Usenet and the Internet, was first published in print 25 years ago. In this piece, we trace the history and impact of the book and of Usenet itself, contextualising them within the contemporary and modern-day scholarship on virtual communities, online culture, and Internet history. We discuss the Net as a tool of empowerment, and touch on the social, technical, and economic issues related to the maintenance of shared network infrastructures and to the preservation and commodification of Usenet archives. Our interview with Ronda Hauben offers a retrospective look at the development of online communities, their impact, and how they are studied. She recounts her own introduction to the online world, as well as the impetus and writing process for \emph{Netizens}. She presents Michael Hauben's conception of ``netizens'' as contributory citizens of the Net (rather than mere users of it) and the ``electronic commons'' they built up, and argues that this collaborative and collectivist model has been overwhelmed and endangered by the privatisation and commercialisation of the Internet and its communities.

\medskip

\noindent
\textbf{Keywords:} Ronda Hauben, Netizens, Usenet, online communities, Michael Hauben, Internet history
\end{abstract}

In the preface to \emph{Netizens: On the History and Impact of Usenet and the Internet}, Michael Hauben defines the \emph{netizen} as something more than just the average Net user: ``people who decide to devote time and effort into making the Net, this new part of our world, a better place''. His choice to combine ``Net'' and ``citizen'' reflected the sense of community and collaboration he found in his interactions on Usenet, the worldwide, decentralised discussion network that predated the Web and was eventually eclipsed by it. The Haubens' convictions underlie the rest of \emph{Netizens}, one of the earliest popular histories of Usenet and its impact.

\section*{The life of the book}

Co-written by Michael and his mother Ronda Hauben, \emph{Netizens} is broken into four sections, beginning with defining and describing Usenet in the mid-1990s (Part I~-- ``The Present: What Has Been Created and How?''), analysing its early origins (Part II~-- ``The Past: Where Has It All Come From?''), and considering the Net's future in an increasingly connected world (Part III~-- ``And the Future?''). The final section shifts focus to technology's role within modern democracy (Part IV~-- ``Contributions Towards Developing a Theoretical Framework''). Throughout the book, the authors emphasise how these new global participatory networks allowed many more individuals access to the public sphere. Through the Net, they argue, netizens gain access to a vast array of information and resources to help them not only be more aware of current events around the globe, but also contribute to materially improving the world. The Haubens' writing process reflected this commitment to resource sharing: the book was published first online in 1994, then in print in 1997 by the IEEE Computer Society Press and later by IEEE-Wiley. Several chapters were also serialised in the \emph{First Monday} journal, founded in 1996 and featuring academic analysis and think pieces about the Internet.

The book's tone and focus were shaped by the authors' own experiences.  Ronda Hauben had been active online since 1988 and was a Usenet regular.  Michael Hauben frequently participated in the Detroit\slash Ann Arbor BBS scene prior to attending Columbia University, where he began researching digital networks' impact. Since the book's publication, Ronda has continued to study how the Internet has empowered citizen journalists, including as a resident correspondent covering the United Nations and UN-related issues. While Michael remained actively involved in discussions both online and offline about the Net's role in participatory democracy, he passed in 2001 by suicide. He had been struck by a taxi cab in 1999, receiving a head injury, and had long struggled with the ensuing complications.

\emph{Netizens} is a much-quoted book~-- alongside the myriad other texts the Haubens produced on the topic of netizens and adjacent issues.  The 1997 IEEE edition, according to Google Scholar, has been referenced almost 900 times. Most notable of these citing works is Howard Rheingold's \emph{Virtual Communities}~(\citeyear{rheingold2000virtual}, revised edition)~-- itself fundamental to the rise of think pieces and ethnographic essays about the Internet and electronic networked media upon its original 1993 publication. Though not an academic analysis \emph{per se}, \emph{Netizens} is widely acknowledged in academic literature in domains such as anthropology, sociology, history, political economy, communication science, cultural and media studies, and journalism studies.

\section*{A wave of Usenet studies}

In the early to mid-1990s, Michael Hauben was studying both computer science and communication, and had professional experience in documentation and information handling. He is representative of a series of technophilic writers, probably best exemplified by Howard Rheingold, who took it upon themselves to write extensively about the new medium they spent so much of their time using, and whose work played a major role in spreading the word about networked social services like Usenet within intellectual spaces.

On the academic side, researchers within the social sciences and humanities were starting to take notice too: in the early 1990s, they started to experiment with the Internet and Usenet as a ground for conducting a new kind of media and cultural studies analysis. Starting with labels such as \emph{cyber-anthropology} or \emph{cyber-sociology}, their discussions led to a new multi-disciplinary domain now known, since the 2000s, as ``Internet studies''~\autocite{wellman2011studying,lueg2003computer}. For instance, Nancy Baym (who later helped found the field of Internet studies and is now a principal researcher at Microsoft) first began analysing Usenet in 1993 while conducting her Ph.D.\ thesis~\autocite[see \emph{inter alia}][]{baym1994from}. There were also student papers, like Bruce Jones's ``An Ethnography of the Usenet Computer Network'' (\citeyear{jones1991ethnography}) and Tim North's ``The Internet and Usenet Global Computer Networks: An Investigation of their Culture and its Effects on New Users'' (\citeyear{north1994internet}).

These very early works were the genesis of research analysing ``cyberculture''~-- as computer-mediated communication and sociality was commonly described during that period. Moreover, much of this work legitimating the Internet and Usenet as worthy fields of study was done by early-career scholars, even undergraduate students. Despite the often-positive tone of the first wave of cyber studies, these early analysts worked at debunking some of the myths about what were then known as ``virtual communities''.

\section*{Usenet between popularity and marginality}

\emph{Netizens} makes a point of featuring Usenet as a burgeoning locus of Internet culture, to the point where it became a metonym for ``the Net'' of the 1990s itself. For a generation of young college students, Usenet served as their first experience with the Internet, and it received substantial attention in English-language mass-market Internet ``how-to'' texts released during this period. Indeed, this period marked Usenet's peak use, followed by a gradual decline at the start of the new millennium. Its remnants and memories have become synonymous with an online ``golden age'', before newbies and spam, and before the Internet had to embrace the ``politics of civilization''~\autocite{fidler2017eternal}.

Thus, Usenet's place in Internet culture is somewhat ambivalent: shifting from an initial position of relative popularity amongst early computer networks users, according to Quarterman's (\citeyear{quarterman1990matrix}) user demographics in the 1980s, to a gradual decline in use at the turn of the new millennium as its underlying communications protocols were superseded by the Internet's TCP/IP~\autocite{russell2014open}. Usenet was used not just for discussion, but also for producing and circulating a huge volume of informational, educational, humorous, and folkloric material, including technical standards, tutorials, jokes, and anecdotes. Much of this output was, at least originally, quite specific to Usenet in terms of content (such as netiquette guides) or form (such as FAQs). In a way, Usenet taught the Internet at large how to communicate and how to behave online.

In a considerable number of countries in Europe, Asia, and Africa, the Usenet ``social'' service (as Quarterman qualified it) and its underlying protocol, UUCP, provided the first digital routes, pathfinders, and map onto which the Internet would graft itself.  Surprisingly, early Unix network administrators have been praised less for their actual first steps in international connectivity (namely, UUCP and Usenet) than for their subsequent role in bridging the gap between their local networking infrastructure and the Internet, which didn't happen until much later in the 1980s, when NSFNET became eager to open trans-international official liaisons (on TCP/IP links). However, this praise sometimes mischaracterises these UUCP\slash Usenet link-ups as the first ``Internet connections''~-- \emph{e.g}., in France, the Netherlands, or South Korea~\autocite[see][]{paloque-bergès2017mapping,paloque-bergès2021real}. This discrepancy can be explained by the strong incentive and will of the Unix community to actually contribute to and use the TCP/IP networks, thanks to a shared computer culture centring on cooperation and friendly rivalry~-- highlighted by the Haubens in \emph{Netizens}. Effectively, UUCP\slash Usenet's international pioneers happened to become, after a few years of experimentation, official ``Internet first-timers'' in their own countries or regions.

But we also have to consider this as a shifting heritage, mostly due to the massive popularity of the Internet beyond the TCP/IP world and the American soft power inscribed and used in the ARPANET\slash Internet filiation~-- as shown, for instance, by ICANN's governance role in software standards~\autocite{nye2010cyber,kalinauskas2013interaction,} and the rise of political issues related to data networks in non-US countries~\autocite{carr2012political,badouard2014internet}.

\section*{Community and networks}

This heritage discrepancy aside, Usenet is also notable for how it bridged the world of computer science with other professional and scientific fields, including early lay users of computer networks, both on- and offline. For example, many early Usenet connections were launched or strengthened at Unix conferences. Usenet can hereby be understood in the long history of community networks~\autocite{schuler1994community}, first established in the 1960s to the 1980s as centralised systems but then later decentralised, including PLATO at the University of Illinois Urbana-Champaign~\autocite{woolley2016plato,rankin2018peoples,}, the public-access Community Memory network in Berkeley~\autocite{felsenstein2016community,farrington1997community} and its homologue in the Netherlands, De Digitale Stadt~\autocite{besselaar2005life}. Such networks must also be located within the broader context of an emerging home computer culture and the associated BBS scene~\autocite{driscoll2022modem}. Within the context of industrialised countries at least, the home computer underwent a seismic transformation throughout the mid-1990s and into the 2000s, shifting from a technical tool and hobbyist passion to something more akin to a middle-class home appliance.

This sense of a digital rights uprising lies at the heart of \emph{Netizens}. Within popular discourse, the Net is presented as an empowering tool for people from marginalised or politically under-represented groups, and many of these users were not only present on Usenet, but also influenced its development, at times inheriting tactics from a longer lineage of offline activism~\autocite{washick2020complaint,mcilwain2019black,kennedy2003gendering,spender1995nattering}. For instance, a fiercely debated proposal for a comp.women newsgroup in 1988 contributed to profound changes in how Usenet functioned and was structured. The discussions and clashes helped raise questions such as, ``What is a newsgroup supposed to be?'', ``What is a newsgroup supposed to want?'', and ``What are community standards?''

The online social networks that have largely supplanted Usenet are nowadays referred to as ``platforms'', but many of their socio-technical features have seen little change since Usenet's heyday~\autocite{weller2016trying} and there remain many barriers to access and use (in terms of economic cost, technological know-how, cultural habits, social structures, etc.).  Today's view of social media tends to be disenchanted. For while most social media platforms have become for-profit, regimental services, their owners are (whether corporate or collective) still grappling with some of the same problems that arose on Usenet, such as acute issues about governance~\autocite{denardis2015internet}, for instance through the prism of ``infrastructure''~\autocite{plantin2018infrastructure}. The complicated politics of netizenship have never been more relevant to broader social and political issues, constrained as they are by structures that are often out of the hands of typical Internet users.

In retrospect, we might hypothesise that access to networked social services, and the capacity to use digital machines and networks easily, if not wisely, was the privilege of a few ``Net elites'': those who knew how the tools (and the workbench, and the whole workshop) actually worked. But another aspect of invisibility is all that has to do with hands-on development and administration of essential network infrastructures and services. Eric S.~Raymond (\citeyear{raymond2020loadsharers}) writes that ``[t]he Internet has a sustainability problem'' because ``[m]any of its critical services depend on the dedication of unpaid volunteers, because they can't be monetized and thus don't have any revenue stream for the maintainers to live on.'' Usenet aficionados, especially those who involved themselves in the daily administration and maintenance of its technical infrastructure, largely behind the scenes, can be seen as (having been) relatively invisible. This issue of invisible labour remains a critical concern, as maintenance of the world's digital infrastructure increasingly falls to un- or underpaid gig and crowdworkers~\autocite[see for instance][]{muntaner2018digital}.

\section*{Cultural and epistemic issues}

The story of Usenet archives is complex, full of bugs and holes, and definitely unfinished~-- it has been documented and criticised by many, Ronda Hauben (\citeyear{hauben2002commodifying}) being one of the first. Usenet archives have made their way to the Web through many layers of so-called ``software as a service'', each layer appropriating the original content and making it more or less accessible~-- though nowadays most such archives are no longer reachable or, at best, in ruins~\autocite{paloque-berges2017usenet}. The argument can be made that ``history will keep in memory what's important'', and while we have seen some Usenet archiving initiatives ``in the wild'' by individuals and companies, it is not altogether clear whether preserving online content for history can rely solely on ``the cloud'', especially if ``the cloud'' is run by commercial concerns to capture and monetise user data.

Usenet archives represent not only a snapshot of time in technological history, but also the lives of their users. The user-run service became quite massive in the 1990s, but to this day, precise demographics and other user data are lacking, and knowledge about users (both qualitative and quantitative) seems to remain a blind spot. Alterations, lacunae, and noise are frequently introduced as part of the archiving process, and all these can significantly impact how the material can be used or interpreted. Digital archives furthermore risk revealing personal opinions and information on topics about which the authors might now feel quite differently.

The study of Usenet and netizens, as subjects of Internet history, addresses the issue of timeliness in research, in particular the perception that our research might already be considered ``dated'' because the technologies studied are considered obsolete in the popular imagination. Alternatively, we wonder what role this research might play as a resource for future historians trying to understand the social and cultural impact of contemporary platforms. For example, some of the only descriptions of content (including excerpts) in identity-focused community fora on proprietary platforms are held in academic research conducted at the time.

In light of all these questions, as well as \emph{Netizens}'s 25th anniversary, we reached out to Ronda Hauben for a retrospective look at the development of online communities, their impact, and how they are studied.\footnote{Explanatory comments in footnotes are those of the interviewers.}

\bigskip

\begin{center}
  * * *
\end{center}

\selectlanguage{american}

\iquestion{Q: How did you first hear about and use Usenet, and in what context?}

I first heard about Usenet at a dinner gathering with users from M‑Net.  M‑Net was an early public-access Unix system set up by Mike Myers in Ann Arbor, Michigan.\footcite[See][3, 5]{mccabe2018history}

At the time (around 1988) I had been teaching an introduction to the BASIC programming language. The class then was called Computer Literacy, but one needed to be able to program in BASIC in order to use the version of the Apple II computer available at the time. The classes were given in what was called the Learning Center set up at the Dearborn Engine Plant of the Ford Motor Company Rouge Complex. I had started teaching at Ford in 1984 but by 1987 executives at Ford appeared to withdraw their support for the computer classes. I remember one incident where visitors to Ford were shown our computer class. The Ford executive showing them around was asked by the visitors what the workers were learning. He responded that they were learning how to type.

At some point I was asked to tell my director at the Center what the curriculum was. I was using a very good book, \emph{A Guide to Programming IBM Personal Computer},\footcite{presley1982guide} teaching how to program in BASIC. I had probably been given that book by the director of the Center to teach when I was hired. With the withdrawal of support for the classes, some decision was made to not announce the computer programming classes as widely as previously. Then classes were canceled under the excuse no one wanted them. Workers and I fought these problems.

Michael was on M‑Net and had been talking about what was happening with the computer classes at Ford. The people he spoke with online were interested in knowing more. Michael showed me how to get on M‑Net so I could share more of what had been happening.

After I had spent time on M‑Net, I appreciated the kinds of communication it made possible. Michael had had several years of experience using hobbyist-run BBSes by the time he found M‑Net. This was my first online experience.

My own computer background by the time I had started teaching at Ford in 1984 was two programming classes, one in BASIC and one in assembler, both at a local community college. Also I had taught at different levels including college, high school, and younger students. I had gotten my Master's degree. By the time I was writing about Unix and Usenet, I had learned some Unix and was using it as I did my research on historical and scientific roots of these splendid new human computer networking contributions to the world. Also I worked as a non-credit student with a professor at the University of Michigan in Dearborn to create a Unix tool to help me with my research. The tool, called signif, searched for patterns in a text.

Our struggle over Ford's efforts to cut out the BASIC computer programming classes eventually gave rise to the \emph{Amateur Computerist} newsletter. The newsletter continues to today.

When I started teaching at Ford, in 1984, Michael was 11 years old. He was quite a computer expert and known on Detroit-area BBSes as ``whiz kid''. Later he changed his handle to ``sentinel''. When I began teaching at Ford, I hired Michael as my consultant and paid him a small fee (\$2 a week) to be available to help me with problems I would run into in my classes with the technology.

By 1987--1988, when the classes were being ended, Michael was 14--15 years old and still my consultant. He also became one of the editors of the \emph{Amateur Computerist} when we started it in 1987--1988. The first issue was February~11, 1988, in celebration of the victory of the Flint Sit-Down Strike on February~11, 1937. (Our publication was begun for and by auto workers, and February~11 marks the date auto workers won the right to unions.)

The struggle we had at Ford with the fight to continue the programming classes attracted interest outside of our immediate situation. For example, Andrew Ross wrote about the struggle over the programming classes at Ford in the book \emph{Technoculture}:\footcite{ross1991hacking}
\begin{quote}
  When worker education classes in computer programming were discontinued by management at the Ford Rouge plant in Dearborn, Michigan, United Auto Workers members began to publish a newsletter called the \emph{Amateur Computerist} to fill the gap. Among the columnists and correspondents in the magazine have been veterans of the Flint sit-down strikes who see a clear historical continuity between the problem of labor organization in the thirties and the problem of automation and deskilling today.  Workers' computer literacy is seen as essential not only to the demystification of the computer and the reskilling of workers, but also to labor's capacity to intervene in decisions about new technologies that might result in shorter hours and thus in ``work efficiency'' rather than worker efficiency.
\end{quote}

The story is also told in the \emph{Amateur Computerist}, Vol.~10 №~2, in Spring 2001.

While I was teaching computer classes at Ford, I learned about MACUL, the Michigan Association for Computer Users in Learning. At its yearly conference there were presentations about teaching about computers. One year I gave a talk about creating a simple BASIC program about the history of computers. (I had used this program in my classes when at Ford.) I was then invited to give my talk at the 1988 yearly conference at ECHO, a similar organization to MACUL in Ohio for teachers of computing. It was during that Ohio conference that I learned about the Cleveland Free-Net\footnote{For more information on this network, see \textcite{grabowski2022cleveland}.} and that it provided free and easy to access to Usenet. But one had to call up a phone connection in Ohio, which was a long-distance charge when calling from Michigan where I was living at the time.

A little later Michael learned he could use the developing NSFNET in Michigan to make a call to the Free-Net in Cleveland without having to pay long distance fees. I learned from Michael how to connect to the Cleveland Free-Net. My first post on Usenet via the Cleveland Free-Net was on January 10, 1992. In the \emph{Netizens} book I describe what happened with my first post to Usenet.

Briefly, I managed to write out a post asking where on Usenet one could have discussions about issues in the history of economics. However the only newsgroup I could find that was even partially related to the topic was misc.books.technical. So I put my post in this newsgroup. Imagine my surprise when I got probably at least 10 responses, basically telling me I had posted in the wrong newsgroup. Several of the posts made suggestions how I could find a more relevant newsgroup, or even create one once I had help from someone who knew how to use Usenet or once I had learned how to create a new newsgroup. I also received encouragement, for example from the netizen who wrote, ``Start discussing on sci.econ. We're all ears.''

When I first got online, I joined sci.econ and posted commenting about issues of the day like large layoffs at General Motors or responses to others' posts. For example, in response to someone else's post suggesting that we propose who should be nominated for the Nobel Prize in Economics, I posted that the creators of Usenet should be awarded the prize as they have created something important for the development of technology and social wealth. During my first year on Usenet, I playfully but also seriously posted that ``reluctantly'' I had decided to be a write-in candidate for the US Presidency since none of the candidates were debating the real issues like we do on Usenet. I didn't win of course, but after the election several people posted raising questions about what the requirements were in different states to be a write-in candidate. There were many other topics I posted on in different newsgroups. Also at the time Michael and I were writing about or giving talks about Usenet and the Internet and we posted related posts or requests for information. Michael for example posted a request for comments in his post ``The Largest Machine: Where Did it Come From and its Importance to Society''. (See Chapter~2 of \emph{Netizens} where I describe his post.) Also in response to a post Michael or I did, one of the ARPANET pioneers sent me a copy of an article by JCR Licklider.  The article by Licklider and Taylor was ``The Computer as a Communication Device''.\footcite{licklider1968a} Then and continuing over the years since, I have found this article of great importance for understanding the vision Licklider and Taylor put forward that inspired others who came after them to create interactive computing based on first the time-sharing of computers and then networking and the Internet. Licklider recognized the computer was not only for number-crunching but, more profoundly, a communication device. In this article he proposes that collaborative modeling is an important capability that the computer will facilitate. I have been doing some talks and writing more recently explaining the significance of this insight that Licklider and Taylor present in this paper. (See for example a paper I wrote in 1998, ``The Internet: A New Communication Paradigm''.\footcite{hauben1998internet})

Also in a more recent talk and draft paper, I explain how this new communications capability has been demonstrated by netizens in South Korea in recent events, even though they may not know Licklider's paper.\footnote{\citeauthor{hauben2017candlelight}, \citeyear{hauben2017candlelight}. A talk based on this article was presented at the re:publica 2017 conference: \url{https://youtu.be/KKFQQDWxNgU}}

\iquestion{Q: What did you know about digital networks when you first got online?}

Of course I knew second-hand about connecting to BBSes using modems from Michael's experience and interest in them. M‑Net was for me a special experience as we had interesting conversations online and even had contact with a staffer to a state representative in the Michigan Legislature in Lansing, Michigan.

When the staffer told us that some legislators were consider taxing online use~-- something that would raise our costs for using M‑Net and other online services~-- we on M‑Net posted our criticism of this proposal. The staffer printed out our criticism and the representative presented it in the committee discussing the proposal. That effectively killed the proposal. This is an example of interesting experiences and discussions I had had with networking prior to getting on Usenet. When Michael and I heard that Usenet was very interesting with many posts, we were eager to find a way to get access.

\iquestion{Q: What kind of computer and computer networking skills did you have, if any?}

By the time I was on Usenet I did have some familiarity using Unix, though I am not sure when I began using it. Later, I even taught one Unix class session at AcIS (Academic Information Systems) at Columbia University when the regular teacher had to be out.

\iquestion{Q: What was your professional status and position when you began writing about Usenet, and what prompted you to do so?}

By the time I had gotten on Usenet, I had had the experience with the exciting computer classes at Ford and that they had been cut out with no input allowed from those taking the classes, nor from those involved in the Learning Center at Ford who knew the response to the classes.

My own professional background was, at that time, that I had taught at several different levels, including two years at colleges. Also, the fight to reinstate the Ford classes continued in various ways which required my studying law and also learning to interact with the political structures. This was also being documented in the \emph{Amateur Computerist} newsletter. I was one of the editors.  Michael's professional experience was with computer support and administration, both at Detroit Mercy College and at Columbia University, as an undergraduate student and after as a graduate student.

Discussions and debates on Usenet and what Michael was learning with his research was very exciting. I shared in this excitement and I was finding what I was learning was very fascinating. (See Chapter~4, ``The World of Usenet''.) I remember discussing how valuable the experience on Usenet was with a Canadian user online. He and I agreed what was happening was something new and important. He didn't think most of those online recognized how important what was happening was. Nor did he understand what made possible this valuable experience. Such conversations helped me to feel it would be good to learn more about how Usenet came to be. And I began to ask for sources to read to understand the development of Usenet. Bruce Jones had written a very interesting paper as part of his graduate work and had begun an archive of helpful material. As far as I remember, at the time Bruce Jones was planning on writing a thesis and/or a book about the story of Usenet's development.  Greg Woodbury's ``Net Cultural Assumptions'' was a fascinating and very helpful article.\footcite[A revised version was later published in \emph{Amateur Computerist} as][]{woodbury1994net} Gene Spafford had written something periodically posted to Usenet in news.misc. There was an article by Henry Spencer and Geoff Collyer, who had written a recent program that was making the scaling of Usenet possible. I even went to Toronto and interviewed Henry Spencer.\footcite{hauben1993interview,hauben1995interview} I was in email contact with some of the pioneers or at least was able to make an email contact with, among others, Stephen Daniel, Tom Truscott, [Mary Ann] Horton,\footnote{Then Mark Horton.} Steve Bellovin, Timothy Murphy, Teus Hagen, Dan Lorenzeni, Piers Lauder, Rob Elz, Dik Winter, and Jim McKie. Bruce Jones's work in setting up the Usenet History Archives and making material available online was especially helpful.

I did online interviews with pioneers like Tom Truscott\footcite{hauben1998interview} and what I learned became the basis for Chapter~10 of \emph{Netizens} (``On the Early Days of Usenet: The Roots of the Cooperative Online Culture'').

\iquestion{Q: How did it become a family affair?}

Michael and I were not only family but also colleagues who helped each other and shared our interest in and support for computer and Internet developments and a number of other common interests. My husband and Michael's father, Jay Hauben, shared these interests as well. At one point Michael said I was one of his best students in terms of netizen developments. He and I were doing related research. We shared what we learned, what we wrote, etc. Another Canadian who Michael met online shared his understanding that there was a need for a collection of articles if there were other articles like what Michael was writing. The tentative title the Canadian netizen proposed for the needed book was ``Readings on the Emergence of a Better World, Due to the Participatory Nature of Public Computer Networks''. I also had a sense independently that there was a need for a book that would include the kind of work Michael and I were doing.

\iquestion{Q: Twenty-five years after the paper publication of \emph{Netizens}, what is the story of the book and your many other writings in the age of emerging networks?}

The first publication of the \emph{Netizens} book was on January~12, 1994, in an online edition. We actually held an event at Henry Ford Community College in Dearborn, Michigan where Jay worked at the time. We invited friends and colleagues and even announced the event in a local entertainment newspaper. As I remember, one person came to the event after seeing that announcement. Others who came were friends or neighbors. At the event, Michael read a passage from one of his chapters and then we went to a room with computers and showed those who attended that the book was online via what was known as a program called ftp (file transfer protocol). The first title of the book was ``The Netizens and the Wonderful World of the Net: An Anthology''. It was also referred to as the ``\emph{Netizens} netbook''. Michael wrote about the event in the Winter\slash Spring 1994 issue of the \emph{Amateur Computerist}:\footcite{hauben1994new}
\begin{quote}
  In honor of the 25th Anniversary of the ARPAnet and of the UNIX operating system and the 15th Anniversary of Usenet News, I am proud to announce a net-book. This net-book provides some of the historical perspective and social context needed to understand the advance represented by the global telecommunications network. The net-book is for those who want to contribute to the care and nurture of the Net.
\end{quote}

His article noted that the book was available via anonymous ftp, gopher,\footcite[First introduced in 1991, gopher is a text-based, menu-driven communication protocol for distributing, searching, and retrieving files and documents via Internet Protocol networks. Though it did have many early adopters, it was eventually replaced by more graphical web browsers like Mosaic. See][]{frana2004before} and at his home page at Columbia University. He asked for comments on the book, explaining that the book was only in draft form at that time. ``We are making it available on-line as we feel it will be helpful for people.  and your comments will help us to make the book more valuable,'' he explained. Also, Michael wrote that ``it would be helpful to have the book published in a printed edition.'' He explained that ``Any suggestions toward this would be appreciated.''

I include Michael's comments in this interview as part of my response about why I was writing about the history and social impact of Usenet.  Both Michael and I were doing so in the context of our efforts to contribute to the Net and the netizens who had access to the Net as well as those who didn't yet have access but who would benefit from knowing why they should find a way to get access. In this context I want to point to Michael's article, ``The Expanding Commonwealth of Learning: Printing and the Net'' (Chapter~16 of \emph{Netizens}).

By Fall 1993 Michael had posted on Usenet an early version of his chapter on the revolution created by the printing press. Michael built on the insights Elizabeth Eisenstein develops in her book, \emph{The Printing Revolution in Early Modern Europe}.\footcite{eisenstein1993printing} He proposed that comparing the two important developments would reveal ``fascinating parallels'' demonstrating ``how the Net is continuing the important social revolution that the printing press had begun.''

Similarly, my chapter on ``Arte'' in the \emph{Netizens} book also takes this kind of broader perspective. Quoting David Hume in ``Of Refinements in the Arts'', I wrote, ``Can we expect, that a government will be well modelled by a people, who know not how to make a spinning-wheel, or to employ a loom to advantage?''

These two chapters were in both the 1994 online version and the 1997 print edition of the \emph{Netizens} book. They express the broad perspective we hoped to encourage in understanding the importance of the Net.

To add to my earlier response to the question about what prompted the \emph{Netizens} book, I want to refer to some of the various forms of encouragement for doing the book that played a role. I previously described the suggestion to Michael from a netizen in Canada who proposed collecting articles about the role the Net can play in the struggle for more grassroots democracy.

Not only was there encouragement but also there was a struggle going on that the book became part of. As an example, I want to describe an event that occurred when I planned to present a talk at MACUL about Usenet.  The talk was accepted for their annual conference in 1993. But when I arrived at the 1993 MACUL conference I noticed a big note posted so all those attending could see it. It listed the presentations in the program that would not be made and were canceled. My presentation on Usenet was in the list of canceled presentations. I tried to find out how that occurred since I had come to the conference prepared to make my presentation. I was told that I had called and said I wouldn't be able to attend. I had done no such thing. The result was that my talk was reinstated and I gave it. Probably some of those who originally planned to attend didn't. But after the conference I was able to have the conference organizers include my presentation in their newsletter to make up for the mistaken cancellation. Then, surprisingly, Michael, who was at Columbia University in NY at the time, was given a copy of my presentation in one of his classes without my name on it, and was told to use it as data in a computer programming assignment for the class.  Michael complained that my name should be put on my article. Later he talked to the instructor who invited me to come to speak with students as a way to make up for the fact of leaving my name off my article. The term was soon over so I wasn't able to get to Columbia University as per the invitation. But I did arrange to go to NY a little later, and on April~24, 1994, the student ACM Chapter sponsored Michael and me to give a program. Students and some faculty attended. Michael and I each made a presentation about an article we had in the \emph{Netizens} netbook. For sometime afterwards, some students would ask Michael if we had found a publisher yet for a print edition of the book. This early edition of the book was followed by an expanded edition online but with a different title, as by then the nature of the Net was changing. So the title was no longer ``Netizens and the Wonderful World of the Net: An Anthology'', but \emph{Netizens: On the History and Impact of Usenet and the Internet}. The process of what was happening is documented in Chapter~12 (``Imminent Death of the Net Predicted'', p.\,214).

That chapter describes the process by which a small group meeting held by invitation only on March~1--3, 1990 at Harvard University implemented an earlier decision to privatize and unleash commercial forces to control the future of the Net. I will take this up further in my response to another question. But first, a few more comments in response to the question about the publication history of the book. We put the first draft online in 1994. Subsequently, a publisher contacted us and offered a print edition. We worked with that publisher until we found he intended to change the content of the book that he did not agree with.  We told him that was not acceptable. He said we would never get the book published in a print edition.

Subsequently, Michael put the table of contents of the book online. We heard from the IEEE Computer Society Press that they were interested in publishing a book about the Net and if we had a manuscript we could send to them. We did have a manuscript. We sent it to them. They put it through their submission process and agreed to publish the book. Also, a publisher in Japan indicated it would be willing to publish a Japanese translation of the book. Michael had some Japanese colleagues, who found translators for the book. So the book was published in a print edition in English in May 1997 by IEEE Computer Society Press and in a Japanese translation in October 1997 by Japanese publisher Chuokoron-Sha. Some time later the IEEE took over publication activity from the IEEE Computer Society Press, with John Wiley and Sons doing the distribution.

Basically the book has had a significant circulation online. And the print edition has been bought and circulated by a number of libraries and other institutions. It is often referred to and Michael's work has especially been referred to. I went to South Korea in 2005 for a conference by the online newspaper \emph{OhmyNews}.\footnote{At ohmynews.com, since 2000.} During the conference there was a visit to the office of the search engine company Naver. The speaker at Naver asked the group from \emph{OhmyNews} for an English word to demonstrate their search engine. I offered the word ``netizens'' in English. The search engine found a number of examples. The representative from \emph{OhmyNews,} who had accompanied us on this field trip said to me, ``You are famous in South Korea.'' Actually, the phenomenon of netizens was famous in South Korea and the same term was even used there, only spelled with the Korean alphabet and with a Korean pronunciation: \begin{CJK}{UTF8}{mj}네티즌\end{CJK} (\emph{netijeun}).

Later, I learned that a South Korean professor had learned of the book from the Japanese version when he visited Japan and another professor then, for a time, used the English version of the book in a one of his university classes. A third professor had written an early book in Korean about Korean networking after reading our book. There are many other interesting examples from around the world of references to the book, with netizens of all ages and occupations building on the work done by Michael and coming up with all sorts of interesting and valuable uses and developments. I have written some articles and given some talks referring to a few of the interesting examples. But it has been a valuable experience finding some of the many references and seeing the range of different contexts and languages, etc. that these have appeared in over the years since Michael first posted his work. Though there have been efforts by those trying to block out the history and experience of the cooperative culture of early networking development and especially of Usenet, there has been great interest in continuing netizen development and learning about and building on Michael's research, for example by researchers from Nigeria, Cameroon, South Korea, Indonesia, China, Philippines, Turkey, Palestine, and the US.

\iquestion{Q: What was the \emph{Amateur Computerist}, home to many of your writings, and how and to whom was it distributed?}

The \emph{Amateur Computerist} started as a continuation of the struggle to get the computer programming classes back that were cut out at Ford.  Some of the Ford workers who had been active in the struggle at Ford, and Michael and I, decided to start a newsletter to keep the fight alive and to continue computer education through the articles in the newsletter. The first proposed name for the newsletter was the ``Beginning Computerist'' but Michael objected to that name and instead proposed the \emph{Amateur Computerist}. One of the United Auto Workers labor pioneers who we had met in another context was willing to write for the newsletter and he was impressed with the name \emph{Amateur Computerist} as for him this referred to someone who did computing for the love of it.

Early on, the newsletter was run off on copy machines and distributed to interested workers at Ford. Later we charged a moderate fee for a year's subscription of four issues by mail, but continued it being distributed for free to interested people at Ford. Eventually we posted it online and also distributed it for free by email to those interested who sent us an email address. In the 1990s, the \emph{Amateur Computerist} was often included on lists of e-zines.

In it we documented the struggle that continued outside Ford to continue the programming classes. We filed complaints with various government officials, contacted newspapers to print articles and printed in the \emph{Amateur Computerist}, for example, the letter we received back from a newspaper editor why no article would appear, etc. We started an alt.amateur.comp Usenet newsgroup and received information from Germany about BBS activity in Berlin and an access-for-all conference in Estonia. Eventually we began publishing articles about Usenet and the Internet in the \emph{Amateur Computerist}. Also, by Fall of 1992 we published a supplement which contained a number of articles about Usenet. The publication of the \emph{Amateur Computerist} has continued regularly for over 30 years after the first issue was published February 11, 1988.

\iquestion{Q: To what extent were you aware of previous publications about Usenet like those of Wendy Grossman (\emph{net.wars}, \citeyear{grossman1997net}) and Howard Rheingold (\emph{Virtual Communities}, 1993), and how did they influence your work?}

I have already listed some of the publications by Usenet users and pioneers that we read. Michael had one of Howard Rheingold's books and valued it. Michael had met Rheingold in 1995 when they were both invited to Japan to give keynote speeches at a conference on netizens and community networks. When I looked online for information about Wendy Grossman's book \emph{net.wars} it indicated it was published in 1997, the same year as the print edition of \emph{Netizens}. I found an online version that I was recently able to refer to. It is of interest, but it seemed to blame the change in culture on something different from what we document in Netizens. We predict in our book that Usenet and the Internet will only continue to thrive if the struggle to protect its public, non-commercial essence is successful.

\iquestion{Q: Let's turn to the topics of democracy and globalization of digital networks, and a reassessment of the netizen\slash ``denizen'' power. We already pre-discussed the idea of \emph{Netizens} as supporting the building of the core representations of digital utopias, especially cyberspace as ``independence'' and/or as another place and ground for citizenship. It might be a simplistic way to formulate how \emph{Netizens} was part of the ``Internet democratic dreams'' of the time, so we would like to discuss this further with you.}

First, I want to present my understanding of the phenomenon of netizens and the nature of the advance represented by Michael's discovery. In 1992--1993 Michael sent out a series of questions online as he was interested in what would be the impact of the Net and how to contribute to it to have a constructive impact.

At the time, Michael was taking a computer ethics class as an undergraduate at Columbia University in New York. The professor had given an assignment to do a research project but not to use books.  Michael raised a series of questions and sent them out broadly online.  At the time there were various networks gatewayed to each other and the Internet, which had been developed over the previous 20 years, and were just becoming more broadly available to those online.

Imagine Michael's surprise when responses came via email from many people explaining how the Net was something important to them and some added that they wanted to help spread the Net broadly and widely.

Michael put together another set of questions and got additional feedback. He wrote an introduction to what would become a paper and posted it online. In it he introduced the phenomenon of the netizen, describing those who had written him.

Eventually the word ``netizen'' and the phenomenon it described began to appear online. And Michael was also getting examples of how it was spreading offline. In the process, in 1995 Michael was invited to give a talk in Japan at a conference about netizens and community networks to be held in Beppu Bay. In his talk, Michael described how two main uses of the word ``netizen'' had developed. One was to describe those who acted as citizens of the Net, which was the use he had intended when he introduced the word. The other was to call everyone online a netizen, which was not the use he intended. The word ``netizen'' in the context Michael used it has nothing to do with ``building core representations of digital utopias''. Michael's use from the very beginning was to describe the netizens who wrote him in response to his questions explaining their actual efforts to use the empowerment the Net made possible. That use was to contribute to the continuing development and spread of the Net so all could have access and help to solve the myriad problems raised in the continuing development of the technology and the social questions that emerge as part of that development.

Chapter~1 in the book, ``The Net and Netizens: The Impact the Net Has on People's Lives'', presents the data Michael based his early research on.  Also, Michael's preface to the \emph{Netizens} book is based on the speech he gave in Japan in 1995 explaining his work.

Essentially the assumption in the question~-- that \emph{Netizens} represents thinking about some utopia~-- is not my understanding, nor was it Michael's. What Michael's work represents is that netizens are a new form of citizenship existing in parallel with the old form of citizenship. So it's not that ``netizens'' is ``dreams'', but rather is reality. Michael discovered the phenomenon and described it, providing the word to represent the phenomenon. Over the years since Michael posted his paper there have been many examples of scholarly papers describing where and how netizens have developed and impacted our society. Perhaps some of the easiest to understand examples have occurred in South Korea. In 2002 netizens elected the President of South Korea Roh Moo-hyun. In 2016--2017 netizens, in massive candlelight demonstrations with whole families participating, won their demand for the impeachment of the president, who then was Park Geun-hye. Between those events there were many other examples of netizen developments.  (This would need an article to adequately describe.) I am focusing on netizens using various forms of online networks, not solely Usenet or something like Usenet. Not only does our book describe the real phenomena of the netizen, but it also presents the broad-ranging vision of pioneers like JCR Licklider, who inspired later pioneers to implement the vision he had presented. Licklider's work continues to provide a broad focus for continuing the development of the Net. Yes, netizens continue to exist in today's online world. When I asked some of those I met in South Korea, ``Are you a netizen?'', the response I often received was, ``I hope so.''

\iquestion{Q: Has the Internet been experiencing a ``tragedy of the digital commons'', and if so, what's your own take on it? In what way do you think Usenet is or was a ``digital commons'', particularly in comparison to today's social media?}

\iquestion{Also, \emph{Netizens} presents an overwhelmingly positive view of Usenet and its core user community.  Despite having been published after the so-called ``Eternal September'', the point when Usenet experienced a sustained flood of new users that overwhelmed its established culture, the book contains very few references to antisocial behaviour such as spamming and trolling. Many people today strongly associate these behaviours with the network and its eventual decline. In retrospect, do you think that \emph{Netizens} was overly optimistic in its assessment of Usenet, its users, and its future?}

In the preface to \emph{Netizens}, actually based on a keynote talk Michael gave in 1995 in Japan, Michael explains,
\begin{quote}
  In conducting research five years ago online to determine people's uses of the global computer communications network, I became aware that there was a new social institution, an ``electronic commons'', developing. It was exciting to explore this new social institution. Others online shared this excitement. I discovered from those who wrote me that the people I was writing about were citizens of the Net, or Netizens.
\end{quote}

Later in the preface, Michael describes how during course work at Columbia University a professor encouraged him to use Usenet and the Internet to do research. Michael explains how his research ``was real participation in the online community, exploring how and why these communication forums functioned.'' Michael elaborates that he ``posed questions on Usenet, mailing lists, and freenets.'' In one of the early chapters of \emph{Netizens}, ``The Social Forces Behind the Development of Usenet'', Michael describes how the ARPANET, CSNET and eventually the NSFNet became involved in providing networking development.

There is in Chapter~10, Appendix~II a list of Usenet newsgroups whose posts were copied from early mailing lists or digests of mailing lists from the ARPANET. They were designated fa.*, which stood for ``from ARPANET'' on Usenet. The list also includes many Usenet newsgroups that were designated net.*. Also in Chapter~10, Steve Bellovin is quoted explaining that ``one of the key elements in the early growth of Usenet was when [Mary Ann] Horton started feeding SF lovers and human-nets mailing lists into newsgroups.'' (p.\,169)

I offer this background to demonstrate the effort both Michael and I made in the chapters that are part of the \emph{Netizens} book to put the networking developments in a context that is broader than merely focusing on Usenet. We did not ignore the role of, for example, mailing lists carried on the ARPANET which contributed to the growth and attractiveness of Usenet when they became available on Usenet. It is important to recognize that Usenet was part of a broader networking culture during the 1980s and 1990s. Some of that culture came into Usenet, for example, from the ability to port mailing lists like human-nets from the ARPANET to Usenet. In the ``Early Days of Usenet'' chapter there is a post describing the importance of participating on a network to be able to understand the nature of human networking. ``By observing what happens when connectivity is provided to a large mass of people in which they can FREELY voice their ideas, doubts, and opinions, a lot of insight is obtained into very important issues of mass intercommunication.'' (p.\,171)

Chapter~10 about Usenet similarly describes how the Usenet users helped to open up communication processes with ARPANET users, as ARPANET was subject to more restrictions than Usenet (pp.\,174--176).

In his talk, ``The Tragedy of the Digital Commons: On the Expropriation and Commodification of Social Cyberspaces'',\footnote{A full recording can be found at \url{https://www.worldsocialism.org/spgb/audio/the-tragedy-of-the-digital-commons/}} Tristan Miller provides the contrast between the open and supportive culture Michael describes that he found on the Net (and especially on Usenet in 1992--1993 doing research for his paper about the developing networks) and what a user will find more recently. If one reads \emph{Netizens}, which was written during and describing the period of the cooperative culture, one will encounter both the background and the challenge to understand what can destroy the cooperative culture.

Through the chapters Michael contributed to the \emph{Netizens} book, he describes the importance of challenging commercialization of the Internet and Usenet. For example, in the preface (p.\,xi) he writes:
\begin{quote}
  But with the increasing commercialization and privatization of the Net, netizenship is being challenged. During such a period it is valuable to look back at the pioneering vision and actions that made the Net possible and examine the lessons they provide. That is what we have tried to do in these chapters.
\end{quote}

In Chapter 1, ``The Net and Netizens: The Impact the Net Has on People's Lives'', Michael writes (p.\,9),
\begin{quote}
  This evidence is exactly why it is a problem for the Net to come under the control of commercial entities. Once commercial interests gain control, the Net will be much less powerful for the ordinary person than it is currently. The interests of commercial entities are different from those of the common person. Those pursuing commercial objectives are only interested in making a profit.
\end{quote}

As an example, Michael describes how CompuServe charged for access by the hour. So people wanting to help others would have a price tag attached. By way of contrast, Michael explains,
\begin{quote}
  The Net had only developed because of the hard work and voluntary dedication of many people. It has grown because the Net is in the control and power of the people at the grassroots level, and because these people developed it. People's posts and contributions to the Net have been the developing forces.
\end{quote}

In \emph{net.wars}, first published in 1997, Wendy Grossman describes a post by Edward Reid which she says was contributed to alt.best.of.internet by Ron Newman. Reid, she explains, demonstrates several AOL software problems that were responsible for the troubles Usenet had with the abuse it received from AOL users. Also, for several years AOL users were charged by the hour for their use of AOL, leading to the problems Michael had predicted would occur if commercial entities got control of the Net.

In Chapter 12 (``Imminent Death of the Net Predicted'') of \emph{Netizens}, there is an account of how the plan to commercialize and privatize the Net was confirmed at an invitation-only meeting at Harvard University held from March 1--3, 1990. Some of the participants were from IBM. MCI and Sprint representatives also attended. The workshop mandate was to set in process a plan to move the Net from a government operation to a commercial service.

An Office of the Inspector General (OIG) report later done about the process of making such a substantial change in NSF policy complained that there was a lack of ``reasoned'' documentation for such a change.  Also, an Acceptable Use Policy (AUP) that had been in place governing how the NSFNET could be used was removed, with no replacement governing how it would continue to protect the public funds used in the NSFNET.

This background I hope points to the commercialization and privatization as a critical aspect of the abuse that developed on Usenet with the use of Usenet by AOL users. Yet this critical aspect of the US government activity which resulted in a fundamental change in the nature of the US portion of the Internet is often left out of or ignored in accounts of how the US portion of the Internet was transformed from a cooperative culture to one dominated by commercial culture and the privatizers' priorities.

I particularly want to commend how Tristan Miller helpfully outlines the distinction between the early cooperative culture and the more recent commercial and privatizers' culture that now dominates the Net. How this change happened is important to understand.

At some point I learned about the \emph{com-priv} mailing list administered by the NSF. I expected to learn about how the decision to commercialize and privatize the NSFNET was made. But there was no such discussion on the list. When I asked, I was told the question had already been decided. Instead, the list was a pro-privatizing discussion. At some point later on, I remember learning that AOL had benefited by having Advanced Network and Services, Inc. (ANS) infrastructure turned over to it, but only a few details were available.\footnote{Advanced Network and Services, Inc. was a US-based non-profit organization formed in 1990 by a trio of NSFNET partner corporations (Merit Network, IBM, and MCI) to maintain the NSFNET Backbone Service's infrastructure. A year later, they also opened a for-profit subsidiary ISP, ANS CO+RE. In 1995, America Online (AOL) acquired this for-profit branch following the decommissioning of the NSFNET Backbone Service.} Basically, there was censorship and secrecy about what was happening to make the changes. And when there was an effort to cancel my talk at MACUL about Usenet on March 12, 1993, I suspected some of those promoting the privatization had a hand in it.

So my sense is that the transformation of Net culture was due to the privatization and commercialization not to particular acts of users. The acts that were a problem were a symptom and made trouble but they weren't the cause.

Also, when I got on Usenet in 1992 it was clear Usenet and the online world it was part of was something special. Both Michael and I realized this and made our efforts to understand the nature of what Michael referred to as a ``new social institution, an electronic commons developing.'' I could only compare it to when I had been in Paris is 1967 and found a very lively and interesting environment at a student area known as the \emph{Cité internationale universitaire de Paris}.

Michael did his research by posing questions and receiving many very interesting responses. I did my research by exploring early online posts and archives, interviewing pioneers, etc. The early cooperative culture still continued when we got on Usenet and for a while afterward, though the coming commercialization and privatization made understanding and knowledge about the cooperative culture all the more important to learn about and understand. The first version of \emph{Netizens} was put online in January 1994. Additional chapters were written and subsequently added. Valuable experiences and discussions continued. See, for example, the discussions about netizen journalism versus the mainstream media in Chapter 13. Also, the US government's National Telecommunications and Information Administration (NTIA) held a virtual conference which Michael and I took part in and each of us did an analysis which we included in the book. (See \emph{Netizens} Chapters 11 and 14,) This online conference made it possible to participate with others throughout the US discussing the impact of the Net and arguing for a future where all would have access.

To answer the question about whether I think \emph{Netizens} was overly optimistic, not at all. Instead, the book did the crucial work of documenting the cooperative culture which was actually created and existed and thus serves as an important model for networking development. As Tom Truscott, one of the creators of Usenet writes in his forward to \emph{Netizens} (p.\,vii),
\begin{quote}
  We learn valuable lessons by trying out new innovations. Neither the original ARPANET nor Usenet would have been commercially viable. Today there are great forces battling to structure and control the information superhighway, and it is invaluable that the Internet and Usenet exist as working models. Without them it would be quite easy to argue that the information superhighway should have a top-down hierarchical command and control structure. After all there are numerous working models for that.
\end{quote}

Actually South Korea and China are two of the countries where netizens have been active in the past few years using the Net to discuss problems and create alternative processes using online discussions for solving the problems. In China, netizens have found ways to create new channels to be able to have an impact on government, while in South Korea an interesting article documented that netizens have developed a culture where there is a new form of citizenship but an old form of government prevails. See ``Analog government, digital citizens'' by Kyung Bae Min in \emph{Global Asia}.\footcite{min2008analog}

It would take additional study on my part to respond to your question about the turning points from the evolution of Internet culture from the collective to the commercial. Several of the events are listed in Chapter 12 of \emph{Netizens}, along with looking at the process by which the NSFNET was commercialized and privatized, as well as looking at how netizens in countries like South Korea have maintained and advanced in their efforts to spread the Net and build offline forms that are based on what they have found possible online. In conclusion for this interview, I want to offer a quote from my abstract for a talk at Stanford University in May 2001 on ``Usenet and the Usenet Archives: the Challenges of Building a Collaborative Technical Community'':
\begin{quote}
  In 1981 [Mary Ann] Horton, one of the early developers of Usenet, wrote that ``USENET exists for and by the users, and should respond to the needs of those users.''

  Almost twenty years later, in the fall of 2000, almost 4000 people signed a petition directed to Deja.com asking them to maintain the archives online that they had compiled of Usenet posts, or transfer it to someone who would continue to keep it online and to provide it with an appropriate home.

  These two events, separated by almost twenty years, help to highlight an important achievement and yet a significant challenge for our times.  Usenet was created as a users' network. What are the implications of this design principle on the continuing development and scaling of Usenet?

  How do the contributions and collaborative efforts by the users affect Usenet's continuing development? The technical collaboration and support that Usenet provides for people around the world is valued, as reflected by the petition to Deja.com. Yet there are problems that develop as Usenet develops, such as the problem of archiving Usenet and maintaining that archive and access to it in a way that recognized the concerns of the online community and provides a means to respond to these concerns.

  As Usenet scales new problems develop. But so too does the body of experience of how to understand and approach these problems.

  Usenet is not only about open source and user-developed content. It is also an example of user involvement in the administration and developing architecture of the network itself. As such, Usenet is a working model of grassroots development.
\end{quote}

That model is a key to what my Canadian friend, and Michael, and I wondered about when I first got on Usenet. I hope the \emph{Netizens} book will continue to help to spread the model and lessons that will make it possible to find ways to build on the model.

\selectlanguage{british}

\printbibliography

\end{document}